\documentclass[9pt,conference]{IEEEtran}
\usepackage{dcase2025}

\usepackage{bm} 
\usepackage{amssymb}
\usepackage{dsfont}

\usepackage{dcase2025,amsmath,graphicx,url,times,booktabs, tabularx}

\title{Self-Guided Target Sound Extraction and Classification \\ Through Universal Sound Separation Model and Multiple Clues}

\name{Younghoo Kwon$^{1}$\sthanks{Equal contribution.},
      Dongheon Lee$^{1, 2*}$,
      Dohwan Kim$^{1}$,
      Jung-Woo Choi$^{1}$\sthanks{Corresponding author.}}
\address{$^{1}$KAIST, School of Electrical Engineering, Daejeon, South Korea \\
$^{2}$Meta Reality Labs, Cambridge, UK\\
}

\begin{document}

\maketitle

\begin{abstract}
This paper introduces a multi-stage self-directed framework designed to address the spatial semantic segmentation of sound scene (S5) task in the DCASE 2025 Task 4 challenge. This framework integrates models focused on three distinct tasks: Universal Sound Separation (USS), Single-label Classification (SC), and Target Sound Extraction (TSE). Initially, USS breaks down a complex audio mixture into separate source waveforms. Each of these separated waveforms is then processed by a SC block, generating two critical pieces of information: the waveform itself and its corresponding class label. These serve as inputs for the TSE stage, which isolates the source that matches this information. Since these inputs are produced within the system, the extraction target is identified autonomously, removing the necessity for external guidance. The extracted waveform can be looped back into the classification task, creating a cycle of iterative refinement that progressively enhances both separability and labeling accuracy. We thus call our framework a multi-stage self-guided system due to these self-contained characteristics. On the official evaluation dataset, the proposed system achieves an 11.00 dB increase in class-aware signal-to-distortion ratio improvement (CA-SDRi) and a 55.8\% accuracy in label prediction, outperforming the ResUNetK baseline by 4.4 dB and 4.3\%, respectively, and achieving first place among all submissions.
\end{abstract}


\section{Introduction}
\label{sec:intro}

The DCASE 2025 Task 4~\cite{yasuda2025}, Spatial Semantic Segmentation of Sound Scenes (S5), aims to detect and separate individual sound events from multi-channel spatial audio inputs. The core objective is to isolate target sound events (foreground sources) from a mixture by distinguishing them from non-target sound events (interference sources) and background noise, and to perform classification. Each audio mixture can contain up to three foreground sources, optionally mixed with interference sources and background noise. Interference sources are differentiated from background noise by their non-diffuse spatial characteristics and their association with specific sound classes. The task defines a set of 18 target classes, while the non-target events encompass a broader set of 94 classes.

The presence of various sound events, including both non-target events and background noise, complicates the application of Universal Sound Separation (USS) techniques. This complexity highlights the importance of Target Sound Extraction (TSE), a method that leverages specific clues to isolate the sound of interest. TSE models leverage various forms of clues, such as a class label~\cite{veluri2023real}, an enrollment sample~\cite{soundbeam, hernandez2024soundbeam}, or a timestamp~\cite{kim2024improving}, to isolate a desired waveform from a mixture. For instance, a class-conditioned TSE model~\cite{veluri2023real} extracts sound events belonging to a specific class. The SoundBeam~\cite{soundbeam} using the enrollment clue can be viewed as an extension of this paradigm, where an enrollment waveform is provided as the clue. The model then extracts sounds from the mixture that match the acoustic characteristics of the enrollment waveform. 

Traditional TSE tasks depend on external cues that are separate from the input mixture. Conversely, in the S5 task, the system must initially detect the target sound events present in the mixture before extracting them. This indicates a self-directed approach: the necessary clues must be extracted from the input mixture itself to guide the retrieval of the target sound events.


\begin{figure*}[h!]
    \centering
    \centerline{\includegraphics[width=18cm]{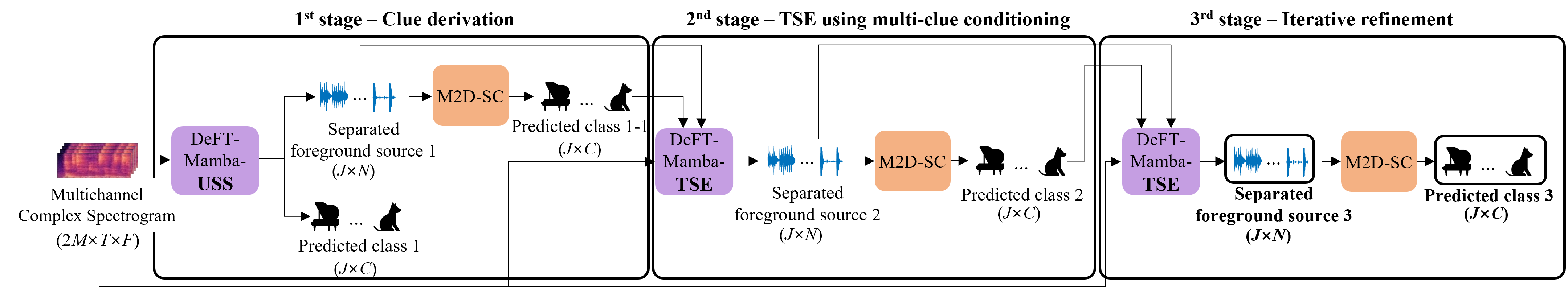}}
    \vspace{-1em}
    \caption{Self-guided multi-stage framework}
    \label{fig:framework}
\end{figure*}

The official DCASE 2025 Task 4 baseline~\cite{nguyen2025} addresses the S5 task by combining Audio Tagging (AT) and TSE. It first performs AT on the mixture using Masked Modeling Duo for Audio Tagging (M2D-AT), a variant of M2D~\cite{hernandez2024soundbeam} specifically fine-tuned for this task. The resulting multi-hot label, representing the predicted classes, is then fed as a clue to a TSE model (ResUNet or ResUNetK)~\cite{kong2023universal} to extract the target events. This approach, however, presents three notable limitations. First, performing AT directly on a complex polyphonic mixture is inherently more difficult than classifying an already isolated sound. Given that mixtures can contain up to six sources, including background noise, accurately identifying all target events is a formidable task. Second, the baseline's AT module cannot leverage the spatial information available in the multi-channel input, which can be critical for disambiguating overlapping sources. Third, the framework lacks a mechanism for iterative refinement; the information flow is unidirectional, and the extracted waveforms are not used to improve the initial predictions.

To overcome these drawbacks, we propose a multi-stage self-guided framework that combines the multi-clue derivation through USS, Single-Label classification (SC), and TSE. The core of our framework leverages a modified version of DeFT-Mamba~\cite{dlee2025}, a state-of-the-art (SOTA) model developed for universal audio separation in multi-channel polyphonic scenarios. This model, which we term DeFT-Mamba-USS, first performs USS to decompose the mixture into estimates for three foreground sources, two interference sources, and background noise. Subsequently, we perform SC on each separated target waveform using Masked Modeling Duo for Single-label Classification (M2D-SC), a version of M2D fine-tuned for classifying the 18 target classes. In the next step, both the separated waveforms (as enrollment clues) and their predicted classes (as class clues) are jointly supplied to DeFT-Mamba-TSE, a variant of DeFT-Mamba adapted for TSE. This multi-clue approach guides the extraction of refined target waveforms. Finally, this process is iterated: the newly extracted waveform is re-classified, and these refined enrollment and class clues are used for another round of TSE, creating a cycle of progressive refinement.

This framework significantly lowers the difficulty of class estimation compared to direct AT on a mixture by performing classification on preliminarily separated sources. The multi-stage refinement process enables the system to achieve progressively more accurate waveforms and corresponding class labels, leading to superior classification accuracy on the evaluation dataset compared to other teams. This approach, which integrates multi-clue injection with iterative refinement, demonstrates great effectiveness, enabling us to achieve the highest position in the competition. Our leading standing has been shown through the class-aware signal-to-distortion ratio improvement (CA-SDRi), a comprehensive metric assessing both separation and classification performance.


\section{Preliminary}



DeFT-Mamba~\cite{dlee2025} is a SOTA model designed to perform USS and classification concurrently. Operating on the complex spectrogram, the model's architecture is composed of $N_b$ stacked blocks of F-Hybrid Mamba and T-Hybrid Mamba modules, which are designed to model relationships along the frequency and time dimensions, respectively. A key distinction from previous speech enhancement models with similar architectures~\cite{wang2021tstnn, lee2023deft, chen2020dual, lee2024deftan} is its replacement of the traditional Feed-Forward Network (FFN) within each transformer block with a Mamba Feed-Forward Network (Mamba-FFN). The model performs separation at the feature level. These separated features are then fed into two parallel decoders: an audio decoder to estimate waveforms and a class decoder to predict their corresponding labels. This dual-head decoder structure effectively resolves the pair-wise ambiguity between the estimated waveforms and their predicted classes, ensuring each separated sound is correctly associated with its label.

\section{Proposed self-guided framework}

The self-guided multi-stage framework performs progressive separation and classification through a combination of USS, SC, and TSE. As illustrated in Fig.~\ref{fig:framework}, our framework consists of three main stages:~\ref{sub:stage1} clue generation via DeFT-Mamba-USS and M2D-SC,~\ref{sub:stage2} TSE guided by multi-clue conditioning, and~\ref{sub:stage3} iterative refinement for enhanced separation and classification.

\subsection{Stage 1: Clue derivation via DeFT-Mamba-USS and M2D-SC}
\label{sub:stage1}
The first stage employs DeFT-Mamba-USS to decompose complex multi-channel spectrograms into distinct object-level features. Unlike conventional architectures of DeFT-Mamba, DeFT-Mamba-USS adopts a modified design with F-Hybrid Mamba and T-Hybrid Mamba blocks. To reduce computational complexity while preserving performance, we exclude the unfold operation and simplify the F-Hybrid Mamba blocks by removing the embedded Mamba modules. This architecture generates six object-level features corresponding to three foreground sources, two interference sources, and one background noise source. Each object feature is then processed by two parallel decoders: an audio decoder reconstructing the waveform, and a class decoder predicting the associated class label. 

\begin{figure}[h!]
  \centering
  \centerline{\includegraphics[width=0.7\columnwidth]{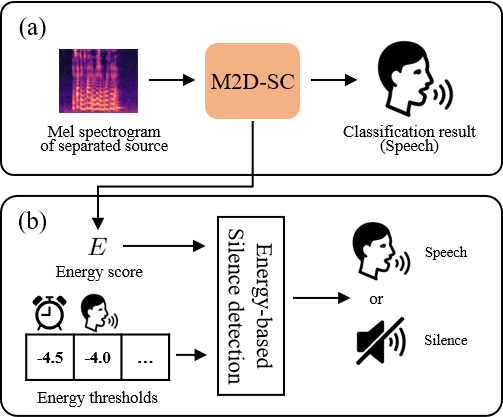}}
  \caption{Inference procedure of M2D-SC (a) The model predicts class and calculates the energy score from unnormalized logits. (b) Silence is determined by comparing the energy score with a class-specific threshold.}
  \label{fig:m2dsc_inference}
\end{figure}

Once the waveform has been reconstructed, each source is fed into a single-label classifier named as M2D-SC. This classifier builds upon the M2D architecture~\cite{niizumi2023masked} and is adjusted specifically for single-label prediction among 18 target classes. Given that some predicted waveforms actually represent silence—indicating non-existent or inactive sources—M2D-SC is also designed to recognize these silences through an energy-based approach~\cite{liu2020energy}. In particular, M2D-SC generates an energy score from its raw logits and employs class-specific thresholds to identify silent segments. As depicted in Figure~\ref{fig:m2dsc_inference}, M2D-SC uses the mel spectrogram from a separated signal as input, and its transformer layers are fine-tuned to predict the signal's class label. The model computes an energy score from the resulting raw logits to determine silence. If this score surpasses a predetermined threshold, the source is categorized as silence irrespective of predicted labels; if not, the model outputs the class assigned by the classifier. The threshold is adapted specifically for each class, as the complexity of identifying silence varies between classes.

\subsection{Stage 2: TSE using multi-clue conditioning}
\label{sub:stage2}
In the second stage, we perform targeted refinement using DeFT-Mamba-TSE, which leverages the clues generated in~\ref{sub:stage1}. DeFT-Mamba-TSE inherits the architectural backbone of DeFT-Mamba-USS but is modified for TSE through multi-clue conditioning. Unlike traditional TSE models~\cite{soundbeam, hernandez2024soundbeam} that encode enrollment clues into embeddings (often resulting in loss of fine-grained details), DeFT-Mamba-TSE injects raw separated waveforms directly. The complex spectrograms of the enrollments are concatenated with those of mixtures along the channel axis prior to the up-convolutional layers of DeFT-Mamba. In parallel, class clues are injected into intermediate feature maps via Residual Feature-wise Linear Modulation (Res-FiLM). Here, class-dependent embeddings $(\beta,~\gamma)$ are computed from the one-hot vectors of the predicted class and consistently applied across all DeFT-Mamba blocks, ensuring strong and stable conditioning throughout the network. After this guided extraction, the refined waveforms are classified again using M2D-SC in the same manner as in the previous stage. This second round of classification not only corrects potential errors from the initial stage but also further refines class clues for the next stage. 

\subsection{Stage 3: Iterative refinement for enhanced separation and classification}
\label{sub:stage3}
In the final stage, we introduce an iterative refinement mechanism to further improve the performance of both separation and classification. The refined waveforms and updated class labels 
are reinjected into DeFT-Mamba-TSE for an additional extraction cycle. This cyclic process allows the system to progressively correct errors and sharpen source boundaries while refining class predictions. At each iteration, new enrollments and class clues are generated internally, making the framework fully self-guided without external supervision. By integrating iterative refinement, the framework effectively mitigates error propagation from earlier stages and achieves superior performance in terms of signal-to-distortion ratio improvement (SDRi) and classification accuracy.

\section{Experimental settings}
The models for three stages were trained individually, and since the output of DeFT-Mamba-USS was used as the enrollment clue for training DeFT-Mamba-TSE, DeFT-Mamba-USS was trained prior to DeFT-Mamba-TSE. To obtain higher-quality speech training data, we replaced the speech data provided from the challenge dataset with the VCTK corpus~\cite{veaux2017vctk} resampled to 32 kHz. In addition, we augmented the percussion class data by collecting additional samples from open-source databases (Pixabay\footnote{\url{https://pixabay.com/sound-effects/}}). These extra sources were spatialized by SpatialScaper~\cite{10446118}, mixing 1–3 target events with a signal-to-noise ratio (SNR) of 5–20 dB and up to two interference events at 0–15\,dB.

\subsection{DeFT-Mamba-USS}
DeFT-Mamba-USS was trained using the same data configuration as the baseline system~\cite{nguyen2025}, and optimized with the AdamW optimizer with a learning rate of 4e-4. The model was trained in a multi-task learning setup, simultaneously performing source separation through the audio decoder and source classification through the class decoder.\\
\textbf{Separation} The negative Source-Aggregated Signal-to-Distortion ratio (SA-SDR) loss~\cite{von2022sa} was applied for estimating foreground and interference sources. Given $M$ estimated signals $\hat{s}_m$ and their corresponding ground truth $s_m$, the negative SA-SDR loss is defined as:
\begin{align}
    \mathcal{L}_{\mathrm{SA-SDR}}=-10\log_{10}{\frac{\sum_{m=1}^{M}\lVert s_m\rVert_2^2}{\sum_{m=1}^{M}\lVert s_m-\hat{s}_{m}\rVert_2^2}} \label{eqn:SA-SDR}.
\end{align}
For the background noise object, the negative Scale-Invariant Signal-to-Noise Ratio (SI-SNR) loss was used:
\begin{align}
    \mathcal{L}_{\mathrm{SI-SNR}}=-10\log_{10}{\frac{\lVert\alpha\cdot n\rVert^2}{\lVert\hat{n}-\alpha\cdot n\rVert^2}},~\alpha=\frac{<\hat{n},n>}{\lVert n\rVert^2} \label{eqn:SI-SNR}
\end{align}
where $n$ and $\hat{n}$ denote the ground truth and estimated background noise. The overall loss for USS $\mathcal{L}_{\mathrm{USS}}$ is formulated as:
\begin{align}
    \mathcal{L}_{\mathrm{USS}}=\mathcal{L}_{F}+\lambda\cdot(\mathcal{L}_{\mathrm{I}}+\mathcal{L}_\mathrm{N})
\end{align}
with the sum of SA-SDR losses for the foreground sources $\mathcal{L}_{\mathrm{F}}$ and the interference sources $\mathcal{L}_{\mathrm{I}}$. $\mathcal{L}_{\mathrm{N}}$ is the SI-SNR loss for estimating the background noise. The losses for interference sources $\mathcal{L}_{\mathrm{I}}$ and background noise $\mathcal{L}_{\mathrm{N}}$ were weighted with $\lambda=0.01$ for concentrating on the separability of the target sound events.\\
\textbf{Classification} The class decoder in DeFT-Mamba-USS predicts the class label for each separated source by minimizing a cross-entropy loss on foreground sources to ensure precise label assignment. For silent or non-existing sources, a Kullback–Leibler (KL) divergence loss was used to enforce the predicted class probabilities to be close to a uniform distribution, thereby avoiding overconfident or spurious predictions on silence. 
\begin{align}
\mathcal{L}_{\mathrm{KL}} = {\mathrm{KL}}\bigl(p_{\mathrm{sc}}\lVert u\bigr)
= \sum_{k=1}^{C} p_{sc}(k) \log \frac{p_{sc}(k)}{u(k)}
\end{align}
where $p_{sc}\in\mathbb{R}^{C}$ denotes the estimated probabilities across $C$ classes for a silent segment, and $u=\frac{1}{C}\mathds{1} \in\mathbb{R}^{C}$ denotes the uniform target for $C$ classes. Additionally, a binary classification branch with a sigmoid output is included to explicitly detect silence, trained using binary cross-entropy loss and thresholded at 0.5 during inference to decide whether a source is active or silent. This combination enables reliable class prediction while robustly handling silent segments.

\subsection{M2D-SC}
M2D-SC is a fine-tuned variant of M2D, and only the last two transformer layers and the classification head were fine-tuned. The M2D-SC was fine-tuned in two steps to maximize classification accuracy while maintaining robust performance for silence detection.\\
\textbf{ArcFace-based Discriminative Training} In the first step, we adopted the ArcFace loss~\cite{Deng_2019_CVPR} to improve inter-class separability and intra-class compactness. For the ground truth class $y_i$ of $i$-th data, the ArcFace loss is defined as:
\begin{align}
    \mathcal{L}_{\mathrm{ArcFace}}=-\log{\frac{e^{s\cdot\cos{(\theta_{y_i}+m)}}}{e^{s\cdot\cos{(\theta_{y_i}+m)}}+\sum_{k\neq {y_i}}e^{s\cdot\cos{(\theta_k)}}}} \label{eqn:arcface}
\end{align}
where $s=32$ is a scale factor, $m=0.5$ is the additive angular margin, and $\theta_{k}$ is the angle between the output feature from the classifier and the trained class center.
For silent segments, KL divergence loss is applied to approximate the estimated probability distribution to a uniform distribution, following the same strategy adopted in the class decoder. \\
\textbf{Energy-based Silence Detection} 
For the energy-based silence detection, we incorporated a hinge loss securing a margin between energy scores of silence and foreground sources. The energy score is given by 
\begin{align}
    E(x)=-\log{\sum_{k=1}^{C}e^{l_k}} \label{eqn:energy_score},
\end{align}
where $l_k$ represents the raw logit value corresponding to class $k$, and $C$ is the total number of classes. A lower energy score indicates a higher likelihood of being an active source, while higher value suggests silence. For the input sample $x$, the hinge loss is defined as
\begin{align}
    \mathcal{L}_{\mathrm{energy}}=\mathbb{E}&_{x_{\mathrm{in}}\sim\mathcal{D}_{\mathrm{in}}^{\mathrm{train}}}(\max{(0,E(x_{\mathrm{in}})-m_{\mathrm{in}})})^2 \nonumber \\
    &+\mathbb{E}_{x_{\mathrm{out}}\sim\mathcal{D}_{\mathrm{out}}^{\mathrm{train}}}(\max{(0,m_{\mathrm{out}}-E(x_{\mathrm{out}}))})^2
\end{align}
where margins $m_{\mathrm{in}}=-6.0$ and $m_{\mathrm{out}}=-1.0$ were chosen to control the decision boundaries. The hinge loss for energy-based silence detection was weighted with a factor of $\lambda_{e}=0.001$. Specifically, The loss functions adapted in each step are as follows: 
\begin{align}
    \mathcal{L}_{\mathrm{SC}}^{1st}&=\mathcal{L}_{\mathrm{ArcFace}}+\mathcal{L}_{\mathrm{KL}}, \\ 
    \mathcal{L}_{\mathrm{SC}}^{2nd}&=\mathcal{L}_{\mathrm{ArcFace}}+\mathcal{L}_{\mathrm{KL}}+\lambda_e\cdot\mathcal{L}_{\mathrm{energy}}.
\end{align}

\subsection{DeFT-Mamba-TSE}
DeFT-Mamba-TSE was trained using the same data configuration as DeFT-Mamba-USS. However, the outputs separated from DeFT-Mamba-USS were used as enrollment clues. This approach ensures that the model focuses on isolating the target source from the mixture rather than replicating the enrollment clue directly. For class clues, ground truth one-hot vectors were employed to minimize confusion and provide explicit conditioning signals. The audio decoder within DeFT-Mamba-TSE was trained using the masked SNR loss~\cite{nguyen2025} to emphasize precise foreground extraction. The masked SNR loss computes the SNR only for active sources, ignoring silent segments.

\section{Evaluation metrics}

\subsection{CA-SDRi}
The official ranking metric of the challenge, CA-SDRi~\cite{yasuda2025}, evaluates both source separation quality and class prediction accuracy by including the SDRi of true positives only. In contrast, false positive or false negative cases act as a penalty by including their numbers in the denominator of the metric. The CA-SDRi is given by
\begin{align}
    \mathrm{CA\text{-}SDRi}=\frac{1}{\lvert \mathcal{C}\cup\hat{\mathcal{C}}\rvert}\sum_{k\in \mathcal{C}\cup\hat{\mathcal{C}}}P_{k}
\end{align}
where 
$\mathcal{C}$ and $\hat{\mathcal{C}}$ denote the sets of ground-truth and predicted classes present in the mixture, respectively. $P_{k}$ is the SDRi of the estimated signals when the class $k$ belongs to $\mathcal{C}\cap\hat{\mathcal{C}}$, and $0$ for the other classes. Silent segments are excluded from this computation, ensuring that the metric focuses solely on active sources.

\subsection{Mixture-level accuracy}
We evaluate the mixture-level accuracy by counting the number of data samples only when the predicted set of labels $\hat{\mathcal{C}}$ exactly matches the ground-truth set $\mathcal{C}$. Writing $\mathds{I}$ for the indicator function and $N$ for the number of data samples, the accuracy is given by
\begin{align}
    \mathrm{Acc}_{\mathrm{mix}}=\frac{1}{N}\sum_{i=1}^{N}\mathds{I}[\hat{\mathcal{C}_i}=\mathcal{C}_i].
\end{align}

\subsection{Source-level accuracy}
Each separated waveform is evaluated independently. Let $M_i$ be the number of separated foreground waveforms from mixture $x_i$. For the the target and predicted labels $y_{ij}$ and $\hat{y}_{ij}$ in the $j$-th waveform separated from $x_i$, the overall ratio of correctly labeled tracks is given by
\begin{align}
    \mathrm{Acc}_{\mathrm{src}}=\frac{\sum_{i=1}^{N}\sum_{j=1}^{M_{i}}\mathds{I}[\hat{y}_{ij}=y_{ij}]}{\sum_{i=1}^{N}M_{i}}.
\end{align}
In the S5 setting, $M_i$ is always 3 because each mixture $i$ contains up to three foreground sources, including those detected as silences. 

\section{Results}
The experimental results are summarized in Table~\ref{tab:eval}. We evaluated five configurations based on different combinations of Foreground Source Separation (FSS) and Class Prediction (CP) available at various stages of the proposed framework. The configurations include (1) \textbf{FSS 1 + CP 1} using the separated waveforms and estimated classes from DeFT-Mamba-USS, (2) \textbf{FSS 1 + CP 1-1} using the waveforms from DeFT-Mamba-USS but processing them by M2D-SC to estimate classes, (3) \textbf{FSS 2 + CP 1-1} performing the second stage processing using DeFT-Mamba-TSE but using the classification results from the first stage M2D-SC, (4) \textbf{FSS 2 + CP 2} using the waveforms separated by DeFT-Mamba-TSE and classes predicted by feeding them into the second-stage M2D-SC, (5) \textbf{FSS 3 + CP 3} applying the two-stage TSE model. Among all configurations, the FSS 3 + CP 3 model achieved the best performance, demonstrating the effectiveness of the proposed two-stage multi-clue framework. Accordingly, this configuration was used to run inference on the private evaluation set for our official challenge submission. These results demonstrate the effectiveness of using USS-derived outputs as multi-clue to perform self-guided target sound extraction.
\begin{table}[h]
    \centering
    \caption{Experimental results of FSS-CP configuration in the proposed framework. CA-SDRi and SNRi in [dB] and $\mathrm{Acc}_{\mathrm{src}}$ in [\%]}
    \label{tab:eval}
    \sisetup{
        detect-weight           = true,
        detect-inline-weight    = math,
        mode                    = text,
        round-mode              = places,
        round-precision         = 1,
        table-number-alignment  = center,
        tight-spacing           = true,
        table-format            = 2.1
    }
    \begin{tabular}{l
                    S[table-format=2.1] 
                    S[table-format=2.1] 
                    S[table-format=2.1]}
        \toprule
        & {\textbf{CA-SDRi} $\uparrow$}
        & {\textbf{SNRi} $\uparrow$} 
        & {$\mathbf{\mathbf{Acc}_{\text{src}}}$ $\uparrow$} \\
        \midrule
        FSS\,1 + CP\,1   & 10.8 & 15.1 & 73.2 \\
        FSS\,1 + CP\,1-1 & 12.7 & 15.1 & 81.8 \\
        FSS\,2 + CP\,1-1 & 14.6 & 18.3 & 81.8 \\
        FSS\,2 + CP\,2   & 14.7 & 18.3 & 83.4 \\
        FSS\,3 + CP\,3   & \bfseries 14.9 & \bfseries 18.4 & \bfseries 84.5 \\
        \bottomrule
    \end{tabular}
\end{table}
Table~\ref{tab:table1} summarizes the results of the leaderboard on the DCASE 2025 Task 4 challenge. The ground-truth annotations are not publicly released for the evaluation set, while the test set includes accessible reference labels. Our system (Rank 1) achieves the highest CA-SDRi on the evaluation set (11.00 dB) and strong mixture-level accuracy $(\mathrm{Acc}_{\mathrm{mix}})$ of 55.80\%. On the test set, it also demonstrates competitive CA-SDRi performance (14.94 dB) and a solid accuracy of 61.80\%. A detailed analysis of these results, along with complete leaderboard rankings and breakdowns, can be found on the official challenge page\footnote{\url{https://dcase.community/challenge2025/task-spatial-semantic-segmentation-of-sound-scenes-results}}.
\begin{table}[h]
\centering
\caption{DCASE 2025 Task 4 leaderboard with CA-SDRi in [dB] and $\mathrm{Acc}_{\mathrm{mix}}$ in [\%].}
\label{tab:table1}
\sisetup{
    reset-text-series = false, 
    text-series-to-math = true, 
    mode=text,
    tight-spacing=true,
    round-mode=places,
    round-precision=2,
    table-format=2.2,
    table-number-alignment=center
}
\begin{tabular}{l*{2}{S[round-precision=2,table-format=2.1]S}}
    \toprule
    & \multicolumn{2}{c}{\textbf{Evaluation Set}} & \multicolumn{2}{c}{\textbf{Test Set}}\\
    \cmidrule(lr){2-3}\cmidrule(lr){4-5}
    Rank & {CA-SDRi $\uparrow$}& {$\mathrm{Acc}_{\mathrm{mix}}$ $\uparrow$} & {CA-SDRi $\uparrow$}& {$\mathrm{Acc}_{\mathrm{mix}}$ $\uparrow$}\\
    \midrule
    1 (Ours) \cite{Choi_2025_t4} & \bfseries 11.00 & 55.80 & 14.94 & 61.80 \\
    2 \cite{Morocutti_2025_t4} &  9.77 & \bfseries 61.60 & \bfseries 15.04 & \bfseries 77.07 \\
    3 \cite{FulinWu_2025_t4} &  9.73 & 51.54 & 14.00 & 59.80 \\
    4 \cite{Qian_2025_t4} &  7.84 & 47.72 & 14.38 & 73.93 \\
    5 \cite{Bando_2025_t4} &  7.55 & 49.51 & 13.31 & 64.07 \\
    6 \cite{Park_2025_t4} &  6.69 & 47.22 & 13.22 & 76.53 \\
    7 \cite{Stergioulis_2025_t4} &  6.60 & 51.48 & 11.12 & 60.67 \\
    8 (Baseline) \cite{nguyen2025} &  6.60 & 51.48 & 11.09 & 59.80 \\
    9 \cite{Zhang_2025_t4} &  3.84 & 22.41 & 11.78 & 65.47 \\
    \bottomrule
\end{tabular}
\end{table}

\section{Conclusion}
We proposed a novel self-guided multi-stage framework for spatial semantic segmentation of sound scenes (S5). By tightly integrating USS (DeFT-Mamba-USS), single-label classification (M2D-SC), and multi-clue TSE (DeFT-Mamba-TSE), the system effectively decomposes complex mixtures and iteratively refines source separation and class prediction. Comprehensive experiments demonstrated that our approach achieves superior performance in both CA-SDRi and classification accuracy compared to existing baselines. The final two stages highlight the effectiveness of using internally generated clues for robust self-guided extraction. These results suggest that leveraging joint separation-classification refinement and multi-clue conditioning can provide a strong foundation for future research in spatial audio scene understanding and beyond.

\section{Acknowledgment}
\label{sec:ack}
This work was supported in part by the National Research Foundation of Korea (NRF) grant funded by the Ministry of Science and ICT (MSIT) of Korean government under Grant RS-2024-00337945, Grant RS-2024-00464269, in part by the BK21 FOUR program through the NRF funded by the Ministry of Education of Korea, in part by the Korean Government (MSIT) under Grant CRC21011, and in part by the Center for Applied Research in Artificial Intelligence (CARAI) funded by DAPA and ADD under Grant UD230017TD.


\clearpage
\bibliographystyle{IEEEtran}
\bibliography{refs}







\end{document}